# 形变碳纳米管选择通过性的分子动力学研究

徐葵，王青松，谭兵，陈明璇，缪灵*，江建军

（华中科技大学电子科学与技术系，武汉 430074）

**摘　要：** 本文采用分子动力学方法，研究了基团修饰后形变碳纳米管的水分子通过性和离子选择性。结果表明，形变碳纳米管的短径与修饰基团的种类、修饰率及修饰位置有关。不同粗细碳纳米管均存在临界短径，小于临界短径的形变碳纳米管具有对氯离子和钠离子的选择性，同时水分子通过速率与本征碳纳米管相比未明显变小。分析系统平均力势表明，离子选择性来源于不同短径碳纳米管管口的通过势垒。对于实际制备中较宽半径分布的碳纳米管，可以通过基团修饰等方法调控其短径，提高其离子选择性。

## 1.引言

碳纳米管自被发现以来，其一维纳米管道输运性质受到广泛关注。Hummer G 等[1][2]通过计算模拟发现，水分子能够高速通过碳纳米管这样的疏水通道，引起了大量实验关注[3][4][5]。当前制造工艺已能够合成碳纳米管阵列构成的纳米级薄膜，并且已经观察到水分子和气体分子通过这些薄膜[6][7][8]。碳纳米管内表面十分光滑[9]和碳纳米管通道内水分子的有序排列[10]，使得碳纳米管内表面与水分子的摩擦很小。通过碳纳米管膜的水分子和气体流量大大超过了根据连续流体力学理论预期的结果，分子动力学模拟中也观察到了这些现象[11]。Su J Y 等人发现碳纳米管通道的水流量在外加电场的作用下会变大[12]。Duan W H 等提出纳米管泵概念，将碳纳米管端口部分作扭转，发现这能提高水分子通过碳纳米管通道的速率[13]。

碳纳米管作为纳米通道不仅有高的水分子通过速率，而且有较好的离子选择性，在海水淡化、污水净化等领域有着巨大的应用潜力。对碳纳米管的计算模拟发现，存在着一个管径区间使水分子能通过而离子不能通过[14]，此时离子通过碳纳米管时遇到的势垒比水分子大得多，以致其较难通过碳纳米管通道。Kevin

*通讯作者 Email: miaoling@hust.edu.cn

Leung 等[15]研究了压力作用下碳纳米管的离子选择特性。在一定压力作用下，碳纳米管能通过水分子，过滤离子，但当压力过于大时，其离子选择性下降。Yuan Q Z 等[16]发现水分子进入碳纳米管通道时，由于极性水分子的有序排列，碳纳米管内部会诱导出电场，对其离子通过性产生影响。对碳纳米管管口[17]及表面[18]的电荷修饰可以加强沿轴向的内部电场，与施加外电场[19]类似，均提高了碳纳米管的离子选择通过性。对碳纳米管的化学修饰同样能改变碳纳米管的离子选择通过性。Gong X J 等[19]在碳纳米管内表面使用羰基氧进行不同模式的修饰，Corry B[20]在碳纳米管顶部端口边缘使用不同化学基团修饰，Chen Q W 等[21]在碳纳米管两端口分别用亲水和疏水基团修饰，使其具有非对称的润湿性。这些对碳纳米管的化学修饰均增强了碳纳米管的离子选择通过性，使一些原本不具有选择通过性的大管径碳纳米管产生了对离子的选择通过性。

碳纳米管的离子选择通过性与其管径大小有着密切关系。因此，提高较宽孔径分布碳纳米管阵列的离子选择性同时不影响水分子的通过速率，对海水淡化、污水净化等应用有很大意义。本文通过基团修饰方法，得到不同短径的形变碳纳米管，研究其离子选择性与短径分布关系。

## 2.模型与计算方法

如图 1 所示，单壁碳纳米管垂直嵌入双层石墨烯片层中构成柱状石墨烯模型（P-G 结构）[22]。对模型三个方向均采取周期性边界条件，使碳纳米管呈六边形阵列排布。其中碳纳米管为一系列不同程度径向形变的（7, 7）（8, 8）（10, 10）（12, 12）扶手椅型单壁碳纳米管，管长均为 1.3 nm。模型原胞大小为 2.46 nm× 2.46 nm×4 nm。如图 1（a），在上侧初始放置浓度为 2.06 M 的 NaCl 海水溶液，其中包含 162 个水分子、6 个 Cl 离子和 6 个 Na 离子。Cl 离子和 Na 离子随机分布于水溶液中。在溶液上方加与碳纳米管垂直的石墨烯片层以模拟静水压的作用，石墨烯片层与碳纳米管上端口的初始距离为 1.1 nm。根据 Zhu F Q 等提出的方法[23]可以求得水分子等受到的静水压为~160 MPa，与类似研究量级相当[20]。水分子诱导电场产生的压强约为~1.83 MPa[16]，远小于静水压。因此在本文计算中只考虑了外加静水压对碳纳米管离子选择性的影响。

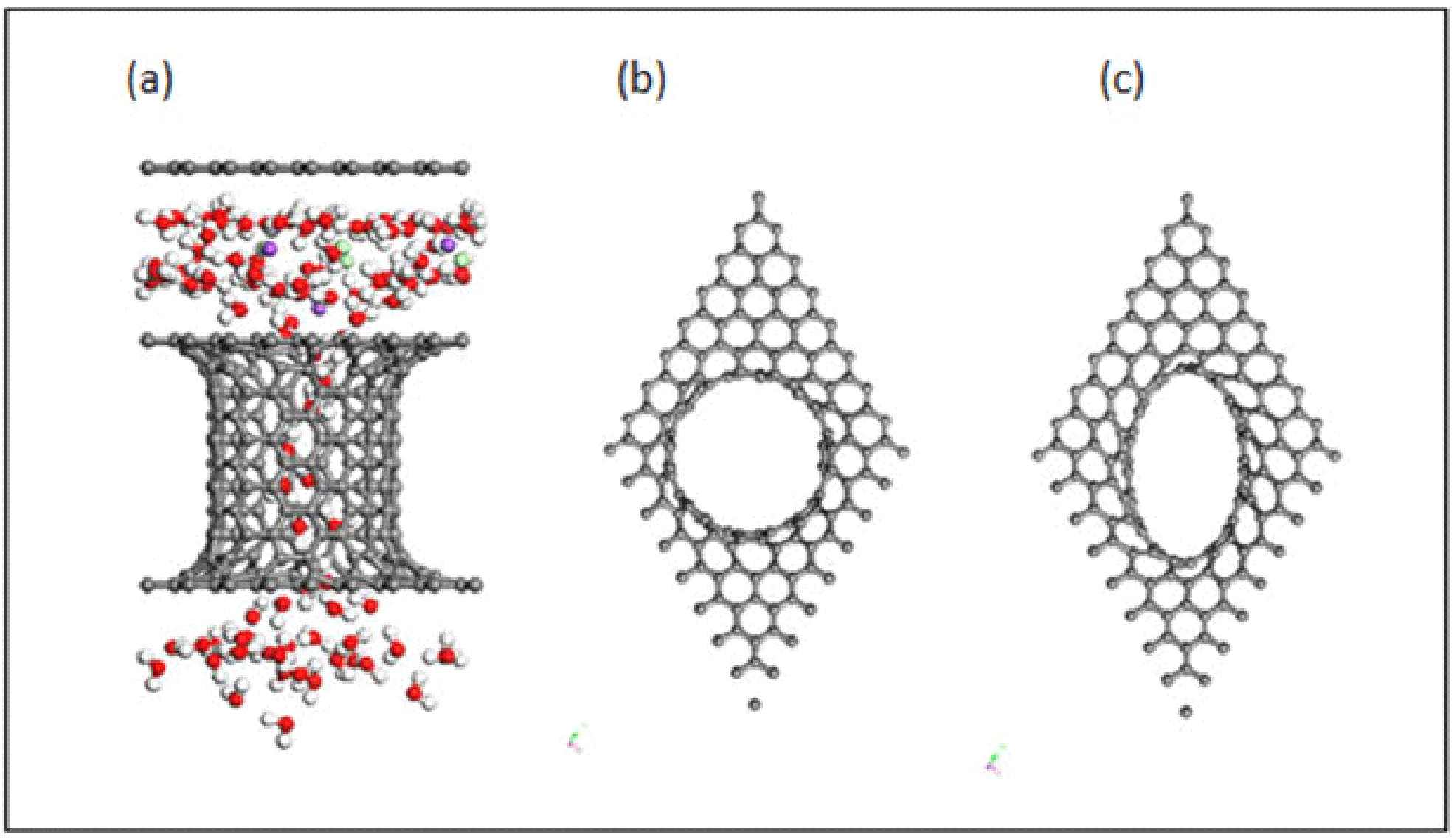

图 1 （a） NaCl 溶液通过柱状石墨烯结构模型侧视图；（b） （8, 8）本征圆形碳纳米管、（c） 短径 0.85 nm 的（8, 8）形变碳纳米管构成的柱状石墨烯模型俯视图

本文采用分子动力学方法[24][25]模拟计算碳纳米管的水分子通过性和离子选择性。模拟中使用正则系综（NVT 系综），体系设置为恒温恒容，采取 Andersen 方法[26]控制体系的热力学温度，维持为 298K。使用 COMPASS 力场[27]对体系的能量进行描述，其中 Lennard-Jones（LJ）势能参数 $r_{i,j}^0$ 和$\varepsilon_{i,j}$ 及电荷量（q）见表 1。采用 Particle-Mesh Ewald 方法[28]来计算系统的静电作用，网格划分精度设置为 0.1 nm。模拟过程中首先进行体系能量最小化弛豫。分子动力学模拟时间步长设置为 1 fs，总时长为 2 ns。每隔 0.5 ns 调节一次石墨烯与碳纳米管的距离，每次向下移动距离为 0.05 nm，以持续提供静水压，保证有效的反渗透作用。

表 1 不同原子的 LJ 势参数和电荷量

| | $r_{i,j}^0$ | $\varepsilon_{i,j}$ | q |
|---|---|---|---|
| C4 | 3.854 | 0.062 | 0 |
| C3a | 3.915 | 0.068 | 0 |
| H | 2.878 | 0.023 | 0.41 |
| O | 3.535 | 0.240 | -0.82 |
| Na | 3.144 | 0.500 | 1.00 |
| Cl | 4.000 | 0.400 | -1.00 |

## 3.结果与讨论

### 3.1 基团修饰对碳纳米管形变的影响

我们采取在碳纳米管边沿使用基团修饰的方法使碳纳米管发生径向形变。修饰基团包括氢原子（−H）、甲基（−$CH_3$）和羟基（−OH）三种，同时考虑了不同

的基团修饰率和位置，共有七种不同修饰模式。故在此有三个因素对碳纳米管形变程度产生影响，分别为修饰基团种类、修饰率及修饰基团排布方式。其中修饰率为单侧修饰基团个数与轴向一列碳原子个数的比值。如图 2 所示为例，使用羟基对碳纳米管两侧进行修饰，此时修饰率为 50%。

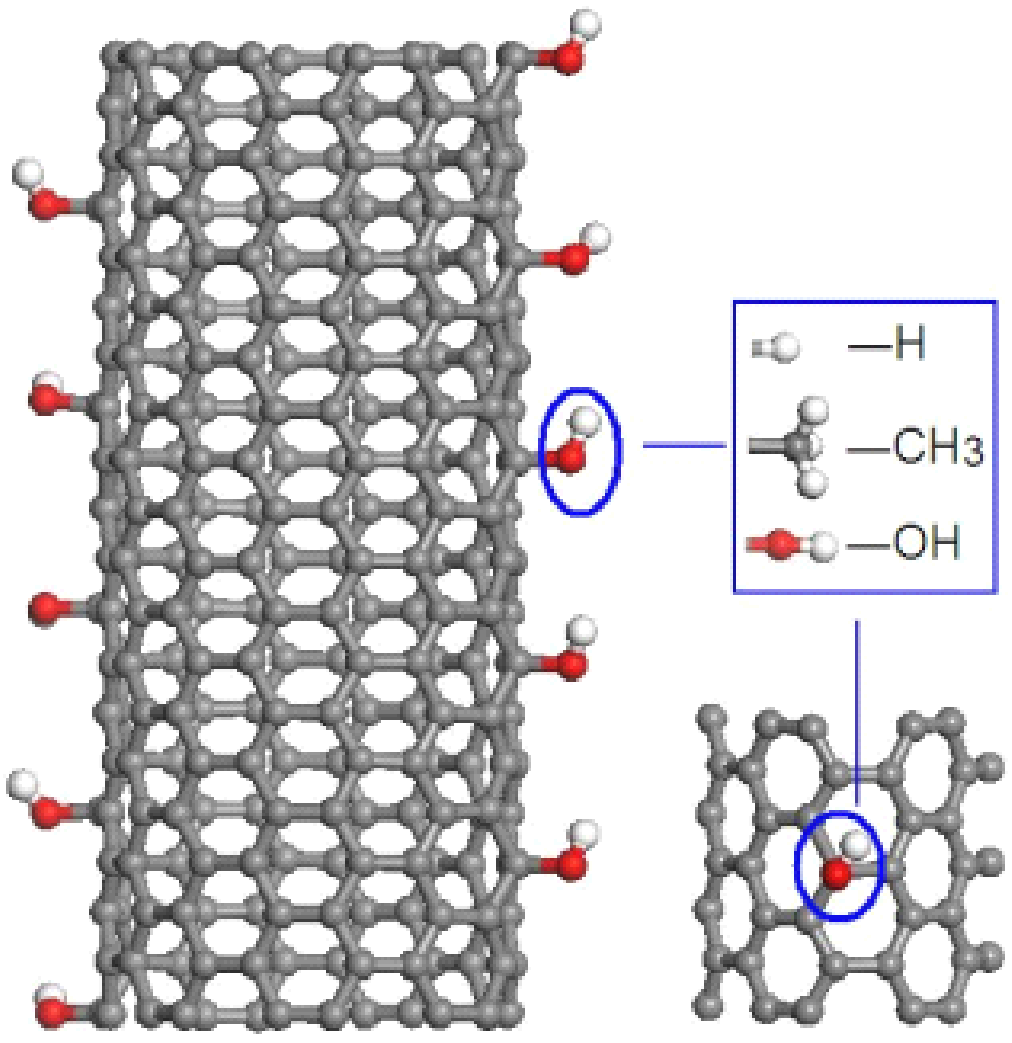

图 2　基团修饰碳纳米管结构示意图

首先对基团修饰后的结构模型进行结构优化。结果显示，对碳纳米管沿着轴向边沿修饰可使碳纳米管从本征圆柱结构形变为椭圆柱结构。其原因为在碳纳米管轴向边沿加上修饰基团后，基团原子与连接处附近的 C 原子形成共价键，C 原子由原来平面 sp2 杂化结构转变为四面体 sp3 杂化结构。在计算中针对 C 原子的两种不同杂化方式 sp2 和 sp3，选取了不同的力场模型，分别为 C3a 和 C4，具体力场参数见表 1。修饰基团与 C 原子间的相互作用，破坏了碳纳米管原有的结构对称性，使碳纳米管沿修饰位置连线方向曲率变大，形成椭圆柱状碳纳米管。

图 3 给出了不同修饰模式对应形变碳纳米管的短径。其中 A、B、C 分别对应三种不同基团排布方式，分别为单排排布、双排排布（修饰原子位置较近）、双排排布（修饰原子位置较远）。图中横坐标为相应的修饰率。可以看出，对于原直径 0.94 nm 的（7, 7）碳纳米管，短径大致在 0.62 ~ 0.82 nm 之间，对于原直径 1.08 nm 的（8, 8）碳纳米管，短径大致在 0.70 ~ 0.98 nm 之间。而且短径的变化与修饰基团种类、修饰率及修饰基团排列都有一定规律。这样对于较宽的短径分布的碳纳米管，意味着可以通过基团修饰方便可控的调整不同碳纳米管的短径，提高管道的离子选择性。

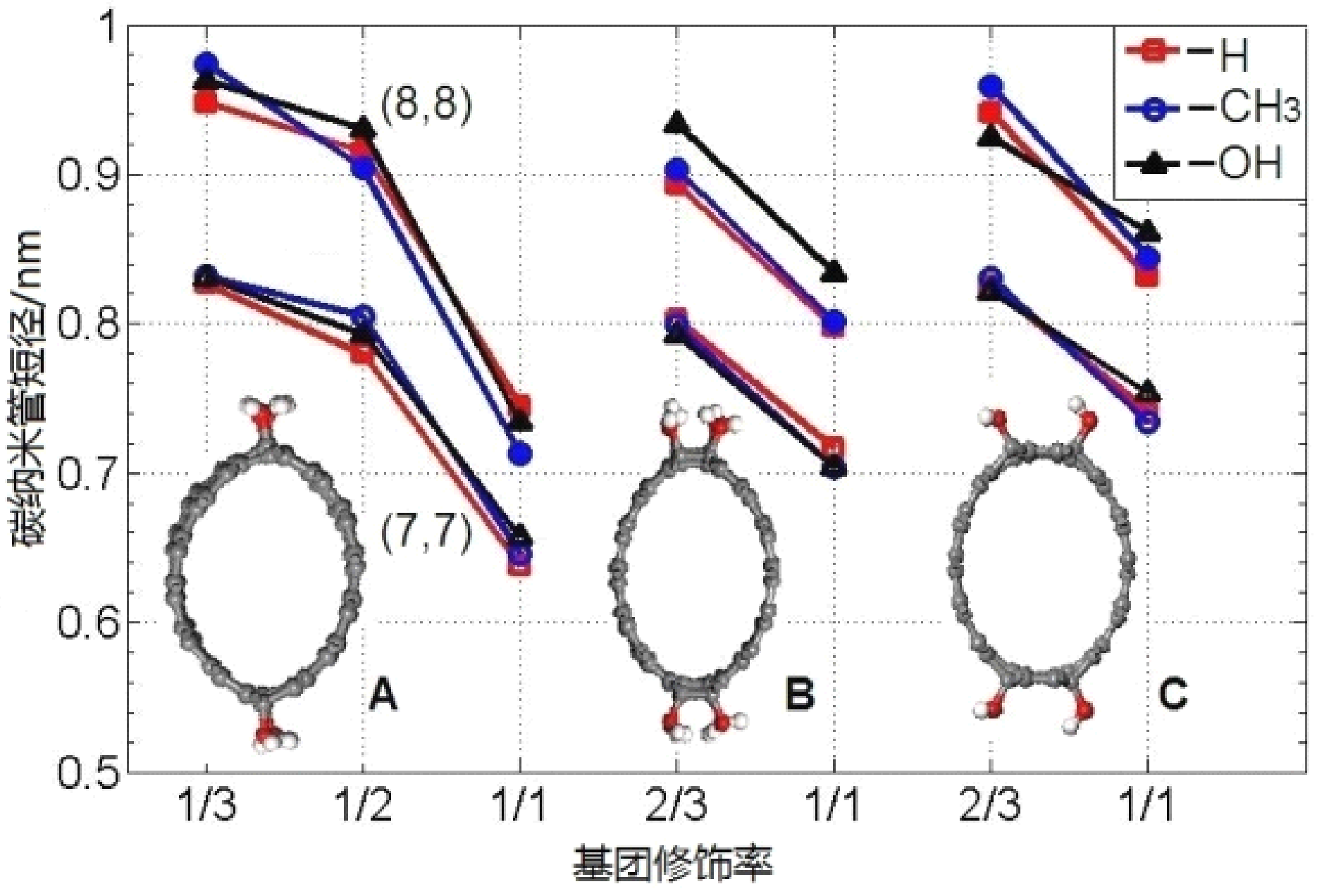

图 3 不同修饰模式对应形变碳纳米管的短径，插图 A、B、C 为不同修饰模式

不同修饰基团对应的碳纳米管短径相差较小，最大仅相差 0.041 nm，即氢原子、羟基及甲基对碳纳米管修饰能力相近。另外不同基团修饰引起的短径差异，在（8, 8）碳纳米管上体现的比（7, 7）碳纳米管上明显。不同基团引起的（8, 8 碳纳米管最大短径差平均值为 0.032 nm,而（7, 7）碳纳米管最大短径差平均值仅为 0.014 nm。这是由于（8, 8）碳纳米管的管径较大，径向的机械强度较低，对不同基团修饰时产生的作用力微小差别更为敏感。

当修饰基团的数目不一样时，碳纳米管径向形变程度差别较大。单排、100%的修饰率修饰时（7, 7）（8, 8）碳纳米管短径与 50%的修饰率相比，平均短径相差高达 0.146 nm 和 0.186 nm。表明修饰基团的数目越多，对碳纳米管的结构作用越强，碳纳米管径向形变越大。

修饰基团位置排布也会对碳纳米管的形变程度产生影响。比较修饰率都为 100%时 A、B、C 三种排布方式修饰后的短径大小，单排修饰后碳纳米管短径最小，随着修饰基团排布分散，碳纳米管短径逐渐变大。说明修饰基团排布越集中，与碳纳米管的作用力方向一致性好，形成的合力大，对碳纳米管径向形变影响大。

## 3.2 形变碳纳米管的水分子通过性和离子选择性

能有效淡化海水的半透膜须具有较大的水分子通过速率和较高的离子选择性。管径较大的（7, 7）及以上的碳纳米管不具有对 Cl 离子和 Na 离子的选择性

[12]。我们选取具有不同形变程度的（7, 7）（8, 8）（10, 10）（12, 12）碳纳米管，研究径向形变后具有较小短径的椭圆柱状碳纳米管的水分子通过性和离子选择性，并与本征圆形碳纳米管进行比较。

比较不同短径碳纳米管的模拟结果，我们得到了（7, 7）（8, 8）（10, 10）（12, 12）碳纳米管对 Cl 离子和 Na 离子产生选择通过性的临界短径。在~160 MPa 静水压作用下，对（7, 7）碳纳米管，当其形变后短径达到约 0.787 nm ~0.81 nm 时，可以完全过滤掉溶液中的 Cl 离子和 Na 离子。对于（8, 8）、（10, 10）、（12, 12）碳纳米管，短径分别须达到~0.83 nm、~0.80 nm，~0.80 nm，就可以对 Cl 离子和 Na 离子产生完全过滤的效果。

图 4 对比显示了 Cl 离子 Na 离子分别在不同短径椭圆柱状（8, 8）碳纳米管（短径 = 0.85 nm，0.75 nm）通过或被阻挡的运动轨迹。由图可见，在模拟时间长度内，Cl 离子、Na 离子却始终无法通过短径为 0.75 nm 的碳纳米管膜，表现出很好的离子选择性。另一方面，Cl 离子、Na 离子通过短径为 0.85 nm 的碳纳米管膜时，在 0.1 ns 左右时进入碳纳米管膜，在 0.25 ns 时通过碳纳米管膜。穿越 1.3 nm 长碳纳米管膜管道整个过程只耗时 0.15 ns 左右，速度约为 8.7 m/s。

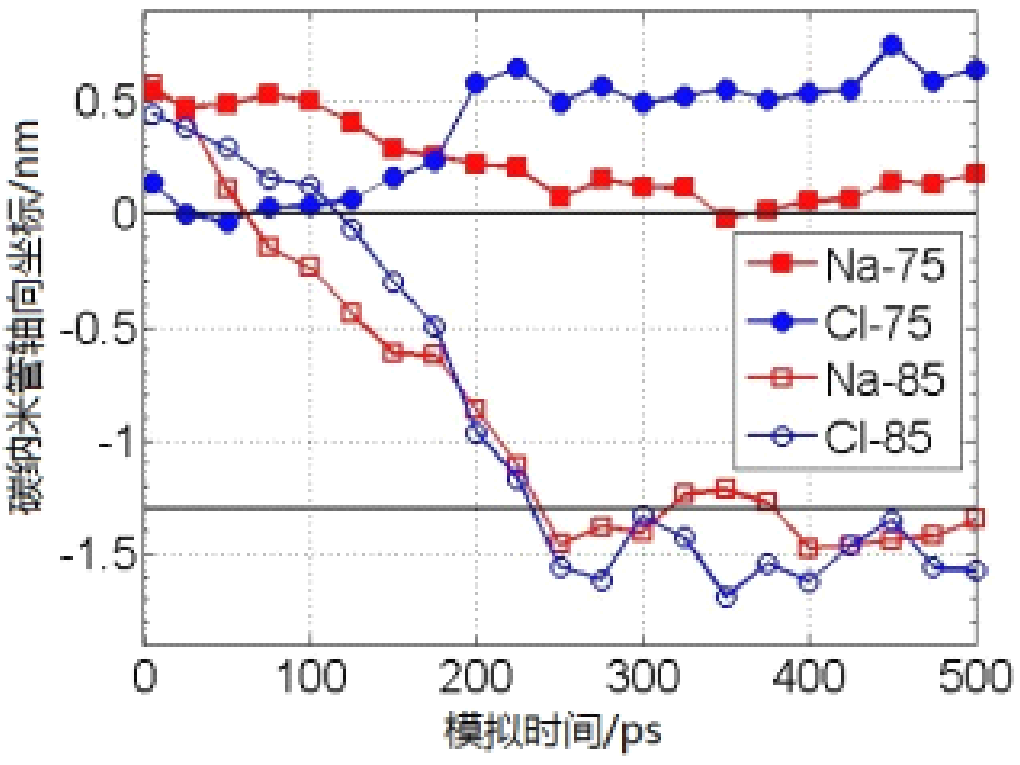

图 4　Cl、Na 离子分别在不同短径椭圆柱状（8, 8）碳纳米管膜（短径 = 0.85 nm，0.75 nm）的运动轨迹。

椭圆柱状碳纳米管中的水分子通过速率是另一个决定其能否真正应用于淡化海水的重要参数。表 2 比较了不同短径（8, 8）碳纳米管中水分子的通过速率。随着短径从 1.077 nm 减小到 0.734 nm，水分子通过速率仅稍有减小，依然高达 23 个/ns。结果表明，对不同短径碳纳米管，整体上均保持着较高的水分子通过速率。这是由于氢键作用，水分子在碳纳米管内部排成规则有序的水分子链，以整体跳动的模式快速地通过碳纳米管通道。水分子链整体的通过，减小了与碳纳

米管内壁的摩擦，使得水分子通过速率得以加快。这与 Alexiadis 和 Kassiinos 计算得到的水分子自扩散系数结果一致[29]。

表 2 不同短径（$D_s$）（8, 8）碳纳米管通道，每 ns 通过水分子的个数（N）

| $D_s$（nm） | 0.734 | 0．758 | 0.774 | 0.795 | 0.815 | 0.835 | 0.853 | 1.077 |
|---|---|---|---|---|---|---|---|---|
| N | 23 | 23 | 18 | 24 | 25 | 28 | 27 | 29 |

对于管径较大的（10, 10）、（12,12）形变碳纳米管，其水分子通过速率相对（8, 8）碳纳米管反而有所下降，不同短径时速率为为 13 到 19 个/ ns。这是由于管径较大的碳纳米管管内能容纳的水分子较多，水分子在其中排布复杂，不能形成稳定的分子链，加大了与碳纳米管壁的摩擦，同时水分子间的粘度也随之增加[30]，导致水分子通过速率降低。

## 3.3 粒子在碳纳米管中能量分布

我们进一步从体系能量的角度来解释前面的模拟结果。Na 离子和 Cl 离子在水溶液中具有水合层,其水合层对离子能否通过碳纳米管通道起重要的作用。当离子直径与碳纳米管管径匹配时，其进入碳纳米管通道也会遇到较大的势垒。此势垒源于离子周围的第一及第二水合层形变及剥离时产生的高能量损失[31]，其大小直接决定了碳纳米管的离子选择通过性。对计算模型中离子周围水分子情况进行统计，发现 Na 离子和 Cl 离子的水分子配位数（氧原子与离子距离小于 0.35nm 的水分子数目[32]）均为 5 个。当通过碳纳米管时，离子受到碳纳米管的作用，其水分子配位数略有变化，并与碳纳米管管径大小直接相关。通过管径为 0.95 nm 的（7, 7）碳纳米管时，Na 离子和 Cl 离子配位水分子维持为 5 个。当（7, 7）碳纳米管形变后短径为 0.83 nm 时，管内部的 Na 离子和 Cl 离子水分子配位数均减少为 4 个，且离子与配位水分子均排布在管道长径方向的平面上。当通过较粗的通道时，如形变后（8, 8）（10, 10）（12, 12）碳纳米管，Na 离子和 Cl 离子水分子配位数保持为 5 个，但离子与配位水分子的排列依然为近似平面排列。离子水合层水分子的剥离及形变导致了碳纳米管管口处的势垒。离子进入较细的碳纳米管通道时遇到的势垒较高，在一定的静水压作用下（本文模拟静水压为~160 MPa），离子也不能进入碳纳米管。由此产生碳纳米管对离子的选择通过性。

图 5 显示了 Na 离子和水分子沿着碳纳米管通道轴向运动时轴向平均力势分布，即体系的自由能分布。由图可知，Na 离子和水分子从管外进入管内时，体系自由能都存在一个阶跃。这意味着 Na 离子和水分子进入碳纳米管通道时会遇到一个势垒。Na 离子在通过短径为 0.75 nm 的（8, 8）碳纳米管通道时，在管口会遇到高达~25 kcal/mol 的势垒，导致其很难通过碳纳米管通道。Na 离子在通过短径为 0.85 nm 的（8, 8）碳纳米管通道时，遇到的势垒仅有~14 kcal/mol。在静水压的作用下，Na 离子还是能通过的。而对于（8, 8）圆碳纳米管通道，Na 离子进入碳纳米管时遇到的势垒很小，几乎可以忽略不计。图 5（b）显示了水分子通过三种碳纳米管通道时的轴向平均力势分布曲线。水分子在通过短径 0.75 nm、0.85 nm 椭圆柱状、本征圆形（8, 8）碳纳米管通道时，遇到的势垒均小于 1 kcal/mol，与 Cl 离子、Na 离子相比非常小。小的势垒使得水分子能以很大的速度通过形变碳纳米管通道，与前面的结果一致。这意味着碳纳米管的径向形变没有对水分子的通过速率造成太大影响，其作为水分子通道仍然非常理想。

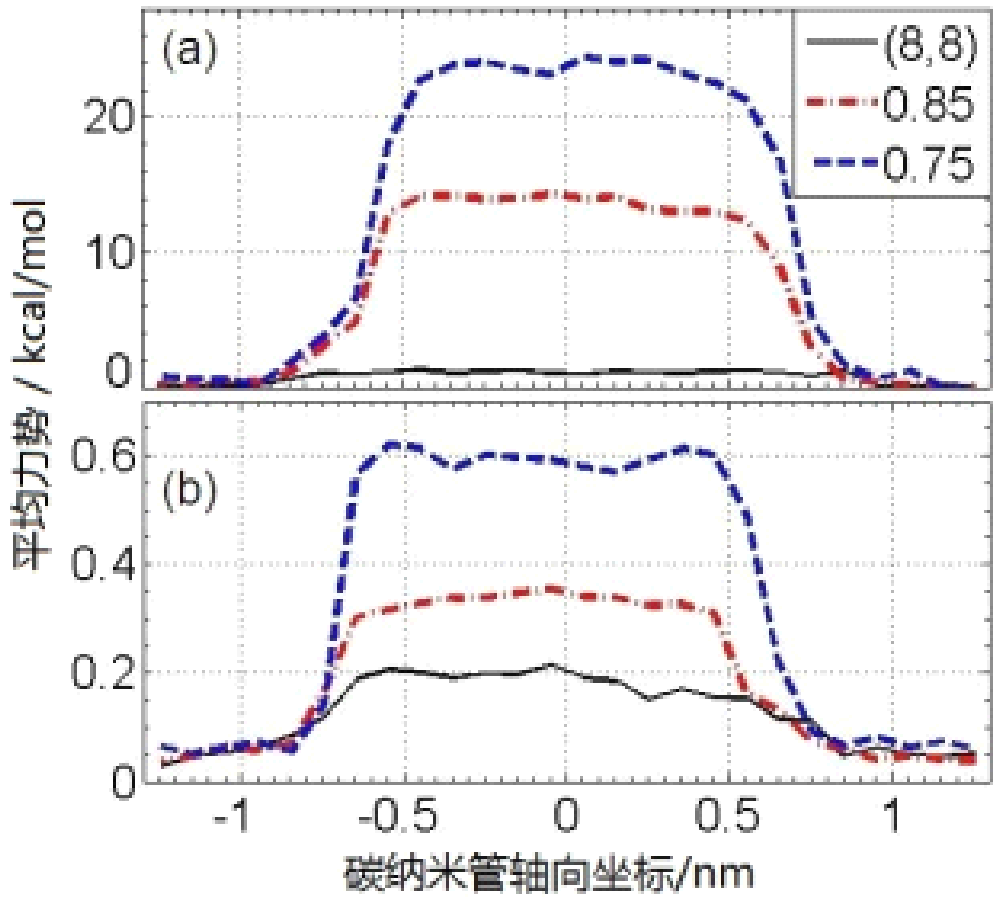

图 5 （a） Na 离子和（b）水分子分别在短径 0.75 nm、0.85 nm 椭圆柱状、本征圆形（8, 8）碳纳米管中的轴向平均力势分布曲线。

## 4.结论

本文对碳纳米管进行基团修饰，使其产生不同程度的径向形变。进一步通过分子动力学方法，对其反渗透过滤盐溶液的过程进行了模拟，研究了形变碳纳米管的水分子通过性和离子选择性。模拟结果表明，在碳纳米管边沿使用基团修饰，能使碳纳米管发生径向形变。碳纳米管形变的程度与修饰基团的种类、修饰率及修饰位置有关。修饰基团越多，位置排布越密，碳纳米管的形变程度越大；而不同基团的修饰对碳纳米管的形变程度影响不大。不同形变碳纳米管均存在临界短

径~ 0.8 nm，小于临界短径的形变碳纳米管具有对 Cl 离子和 Na 离子的选择通过性，同时水分子通过速率与本征碳纳米管相比未明显变小，仍然有着很高的水分子通过速率。对平均力势的分析表明，离子通过碳纳米管时，在管口会遇到一个与管短径有关的势垒。当势垒高到一定程度时，离子将不能进入碳纳米管。碳纳米管对离子选择通过性正是来源于于此。

对于实际制备的碳纳米管，其管径分布较宽，而碳纳米管的离子选择性与其管径密切相关。我们的研究表明，可以通过基团修饰等方法调控碳纳米管的管径，使其离子选择性得到提高。碳纳米管实际用于淡化海水会更加可行。我们的研究对未来碳纳米管在海水淡化领域中的发展，以及使碳纳米管作为解决水资源短缺问题的方案具有一定的参考意义。

## 参考文献

# molecular dynamic of selectivity and permeation based on deformed carbon nanotube

Xu Kui, Wang Qing-Song, Tan Bin, Chen Ming-Xuan, Miao Ling, Jiang Jian-Jun

*(Department of Electronic Science and Technology, Huazhong University of Science and Technology, Wuhan 430074, China)*

**Abstract:** Extensive molecular dynamics simulations of water permeation and ion selectivity of the single-walled carbon nanotubes with the radial deformation are presented . The simulated results indicate that there is a close relationship between the minor axis of deformed carbon nanotubes and the variety, density as well as the position of functional groups. The critical minor axis of different diameter carbon nanotubes exists, and the carbon nanotube whose minor axis is less than the critical minor axis owns the selectivity of chlorine and sodium ions. Meanwhile, compared with intrinsic carbon nanotubes, the deformed nanotubes have not obviously decreased the permeation of water. The analysis to the potential of mean force reveals that the selectivity and permeation of ions comes from the pass potential barrier of carbon nanotubes with various minor axises. Furthermore, our observations of modifying with functional groups may have significance for controlling the minor axis and improving the selectivity and permeation of ions, when it comes to some large nanotube in real manufacture.